\documentstyle[twocolumn,aps]{revtex}
\begin{document}

%
%

\title{K S Krishnan and the early experimental evidences for
the Jahn-Teller Theorem}

\author{G. Baskaran}

\address{ The Institute of Mathematical Sciences\\
    C.I.T. Campus\\
       Madras 600 113, India}

\maketitle

\begin{abstract}

Jahn-Teller theorem, proposed in 1937, predicts a distortional 
instability for a molecule that has symmetry based degenerate
electronic states.
In 1939 Krishnan emphasized the importance of this theorem for
the arrangement of water molecules around the transition metal or 
rare earth ions in aqueous solutions and hydrated saltes, in a short
and interesting  paper published in Nature by pointing out atleast 
four existing experimental results in support of the theorem. 
This paper of Krishnan has remained essentially
unknown to the practitioners of Jahn-Teller effect, eventhough
it pointed to the best experimental results that were available,
in the 30's and 40's, in support of Jahn-Teller theorem. Some of
the modern day experiments are also in conformity with some
specific suggestions of Krishnan.

\end{abstract}

Jahn-Teller effect$^1$  is a beautiful and simple quantum phenomenon
that occurs in molecules, transition metal complexes as well as solids 
containing transition metal or rare earth ions.  It states roughly that 
`a localized electronic system  that has a symmetry based
orbital degeneracy, will tend to lift the degeneracy
by a distortion that results in the reduction of the symmetry on 
which the degeneracy is based.' 
In isolated systems such as a molecule or a transition metal complex it is 
a dynamical or quasi static phenomenon.  They are called dynamic
and static Jahn-Teller effects$^2$. When it occurs co-operatively 
in crystals it is a spontaneous symmetry breaking phenomenon and a 
crystal structure change.  This is called a co-operative 
Jahn-Teller effect.

Even before Jahn-Teller theorem appeared, 
Krishnan and collaborators$^3$  performed a  series of pioneering magneto 
crystalline anisotropy study of families of paramagnetic salts 
containing transition metal and rare earth ions lending  good support 
to various new quantum mechanical ideas including those of Bethe, 
Kramers and Van Vleck on crystal field splitting and magneto crystalline 
anisotropy. In the biographical Memoirs of the Royal Society of
London,  K. Lonsdale and H.J. Bhaba$^4$ wrote: `The papers published by 
Krishnan during this 
period (30's) in collaboration with B C Guha, S Banerjee and N C Chakravarty
were the foundation stones of the modern fields of crystal magnetism
and magnetochemistry'.  

In the paramagnetic salts that Krishnan and collaborators studied
the magnetic ions are well separated from each other by the intervening 
water molecules
and also anion groups.  As a result any direct or superexchange or
dipolar interactions between the magnetic moments are weak;
consequently any co-operative magnetic order such as antiferrmomagnetic
order is pushed down to very low temperatures below 1 K.  This 
enabled Krishnan  and others to study in detail the magnetic 
properties of essentially isolated paramagnetic ions in various 
crystal field environments. 

Jahn-Teller theorem and its experimental consequences has been 
studied  in great detail in chemistry and physics particularly
in the context of electron spin resonance experiments$^5$.  
In the simplest of transition metal complexes there is a cubic 
(octahedral) environment around the transition metal ion such as $Cu^{2+}$.  
The octahedral environment leads to a crystal field splitting of the 
five fold degenerate d level into an orbital doublet ($e_g$) and a 
triplet ($t_{2g}$).  

In Krishnan  and collaborator's work one 
notices repeated reference to deviation from cubic electric 
field at the center of rare earth ions in several cases and
transition metal ions in some cases
as inferred from their own anisotropic magnetic susceptibility
measurements.  The most obvious causes
for departure from regular cubic symmetry of the coordination 
clusters are inequivalence of the ligand atoms in the first or
second co-ordination shell and forces of crystal packing.  
No one suspected that electronic orbital degeneracy can lead to 
an asymmetry such as a distortion of the octahedra with 
equivalent ligand atoms.

It is at this juncture the theoretical papers of Jahn and Teller
appeared, which suggested another important cause for molecular
asymmetry.  And Krishnan readily appreciated the importance of 
this theorem for crystal field splitting and arrangement of water
molecules around paramagnetic ions  in aqueous solutions   
and wrote an interesting  short paper in Nature$^6$ in 1939
that we reproduce here. It is interesting that Van Vleck$^7$, 
who very much admired$^8$  K S Krishnan, also developed his
theory of Jahn-Teller effect for orbital doublets in paramagnetic
ions in the same year.

In his paper KSK quotes at
least four existing experimental results that support  
the Jahn-Teller theorem:  i) x-ray data that is consistant with  
a small deviation of the perfect $H_2O$ octahedra around  
the paramagnetic ion in hydrated salts
ii) magnetic data, mostly from his group, that exhibits strong
magnetic anisotropy that is similar in magnitudes in various salts
suggesting that the cause for any distortion arises from the electronic
state of the paramagnetic ions rather than the surrounding atoms 
iii) asymmetry inferred from electronic absorption spectra of
cation surrounded by water molecules in aqueous solutions
studied by Freed et al. 
iv) magnetic double refraction exhibited by the aqueous solution
of these salts, as observed experimentally by Raman and
Chinchalkar and Haenny and v)  the experimental observation
of Chinchalkar that double refraction is absent experimentally
when the paramagnetic ion is in an S-state (for example $Gd^{3+}$ 
or $Mn^{2+}$ which are orbital singlets)

Krishnan's paper contained important suggestions and also looked
at a slightly more complex system namely rare earth ions in 
aqueous solutions than the relatively simpler case of transition metal ions 
in solids. What Krishnan was looking for was perhaps static Jahn-Teller
distortion at room temperature.  The issue of timescale
associated with the distortion dynamics and the nature of the 
experimental probe becomes important here. 
The Jahn-Teller cluster being a finite system, will either 
quantum mechanically tunnel among equivalent distorted configurations 
in a phase coherent manner with a short time period , or will hop to 
various equivalent distorted configurations incoherently 
through thermal fluctuations. Optical absorption, for example, is a 
short time scale measurement: it  will see  even dynamic distortions 
as a static one.  The static susceptibility measurememt on the
other hand is a low frequency probe.  Any indication of distortion
through this experiment will mean a nearly frozen distortion.

The choice of paramagnetic salts dissolved in water offers some
special advantage too. A paramagnetic ion is typically surrounded by
six water molecules forming a rigid octahedral complex.  
A coordination complex such as  $[Cu (H_2 O)_6]^{2+}$, is loosely coupled to
the environment namely water. This makes the restoring force 
against the Jahn-Teller distortion weaker, making a nearly static
Jahn-Teller distortion feasible even at room temperatures. 
On the other hand, In the case of a solids,  the Jahn-Teller 
distortion has to work against the packing forces of the crystal 
thereby reducing the Jahn-Teller stabilisation energy.

Many of Krishnan's experiments involved rare earth ions.
In view of the larger degeneracy and also the compactness of
the f-orbitals, rare earth ions have some advantages as well
as disadvantages.  The compactness of the wave functions make
the crystal field splitting smaller and also the quenching of
angular momentum less important.  On the other hand the
spin orbit coupling, which can make the magnetic anisotropy
stronger, is larger in the rare earth case.  The larger degeneracy
of the f-orbitals also leads to a proliferation of low lying multiplets
making the theory as well the interpretations of the spectra hard.
Perhaps because of these difficulties most of the classic studies 
of Jahn Teller effect seems to be confined to transition metal ion systems.

I will try to put some of Krishnan's suggestions in the modern 
context.  Now it is well known  that the dynamic to 
static Jahn-Teller distortion cross over temperature, as measured by ESR
measurements, ranges from 1K to
more than 300K in a variety of systems$^{10}$. Some of these room 
temperature systems include $Cu^{2+}$ ion that
Krishnan was referring to based preseumably on room temperature
measurements.  What I mean to say is that static Jahn-Teller 
distortion is not a very low temperature luxury that Krishan could 
have missed.  Secondly, Krishnan
was referring to some x-ray structural data as evidence for octahedral
distortion.  some modern EXAFS measurements$^{11}$ on $Cu^{2+}$ and 
$Cr^{2+}$ ions in aqueous solutions at room temperatures show the 
presence of a distorted octahedra with two copper-oxygen distances 
of about 2.00 Au and 2.3 Au.  

Krishnan's suggestion that the large magnetic double refraction
observed by Raman and Chinchalkar and Haenny in aqueous solutions of 
paramagnetic ions is very interesting.  This seems to be a compelling 
evidence for the Jahn-Teller distortion, as explained by Krishnan:
the distorted octahedral complex, inview of the magnetic anisotropy 
arising from the large spin-orbit coupling, tend to align themselves 
in water along the magnetic fields causing double refraction.  
The absence of solid state effects and diluteness of the paramagnetc 
ions makes this argument very important: the system resembles somewhat 
an ideal gas of `octahedral molecules' as far as the double refraction
is concerned.    I should also point out that I have not been able to 
find out in my short literature survey any later study of Jahn-Teller 
effect in aqueous solutions using the magnetic double refraction
as a proble.

After this one paper on Jahn-Teller effect, Krishnan essentially 
left the field of crystal paramagnetism. Further he also never wrote 
any paper on the issue of Jahn-Teller effect.   

The insightful suggestions of Krishnan, that
came soon after the Jahn-Teller theorem,  was never taken 
up for further study as far as I can see in the literature.  
This paper  has remained unknown.  One wonders why.  
One possible reason is the choice of rare earth ions, which as
we mentioned earlier is complicated theoretically and also 
hard when it comes to interpretation of the experimental
spectra.  Another possible non scientific reason is as follows:
a  detailed followup paper or some further works by Krishnan$^9$
could have made Krishnan's work known more and opened up the 
field of Jahn-Teller effect much earlier,
making Krishnan a pioneer in one more beautiful field.

Detailed  experimental confirmation of Jahn-Teller theorem became
possible only after the ESR experiments of Bleaney and Bowers$^{12}$
in 1952 in $CuSiF_6.6H_2O$ and related family of paramagnetic salts, 
about 15 years after the enunciation of the theorem
and Krishnan`s supporting suggestions. 
The milestones included careful observation of the split ESR line
below about 50K (indicating frozen distortion in the ESR time scales)
and motional narrowed unsplit line above about 50K (indicating a dynamic
distortion) for the copper salts. 

Narly half a decade of experimental work on 
Jahn-Teller effect in the modern era is broadly consistant 
with the sixty year old suggestion of Krishnan. 
The practically unknown suggestions of Krishnan , however,
remains as the only experimental support for the Jahn-Teller 
theorem covering a period of more than ten years,
before the beginning of the modern era.  As I mentioned earlier,
the suggestion of magnetic double refraction by Krishnan
may be still be useful in the modern context for Jahn-Teller
effect study in the rich variety of new and old coordination complexes
involving transition metal and rare earth metal ions.

{\bf Acknowledgement}

I wish to thank Dr Tapan Kumar Kundu (RSIC, IIT Madras) 
for an informative discussion.

\vspace{0.5cm}
{\bf References}\\
\vspace{0.2cm}
1) H.A. Jahn and E. Teller, Proc. Roy. Soc. {\bf A161} 220 (1937)\\
2) M.D. Sturge, in  Solid State Physics,  Eds. Seitz and Turnbull,
vol. {\bf 20} page 91 (Academic Press, New Yor, 1967);
I.B. Bersuker, Jahn-Teller effect and vibronic interactions
in Modern Chemistry (Plenum, NY, 1984)\\
3) Starting from 1933 Krishnan and collaborators wrote more than a dozen
important papers on exhaustive study of magneto crystalline anisotropy in
paramagnetic salts\\
4) K Lonsdale and H J Bhabha in Biographical Memoirs of the 
Royal Society of London, 1962\\
5) A. Abragam and B. Bleaney, Electron Paramagnetic Rosonance of
transition metal ions (Clarendon Press, Oxford 1970)\\
6) K. S. Krishnan, Nature, {\bf 143} 600 (1939)\\
7) J. H. Van Vleck, J. Chem. Phys. {\bf 7} 72 (1039)\\
8) P.W. Anderson, in his K.S. Krishnan Memorian Lecture devivered
at NPL, Delhi in 12th January 1987, recalls warmly Van Vleck's
(Ph.D. supervisor of P W Anderson)
admiration for K.S. Krishnan.  Another 
such instance involving Van Vleck is quoted in the article
by K.Lonsdale and H.J. Bhaba (ref. 4)\\ 
9) A close look at K.S. Krishnan's style of paper writing shows
that typically he starts with a short letter style paper and ends
with a series of long papers presenting the details of his 
experimental investigations.  Roughly in each five, six  years he
goes for new pastures.\\
10) T.K. Kundu and P.T. Manoharan, Chem. Phys. Lett. {\bf 241} 627
(1997); {\bf 264} 338 (1997); T.K. Kundu, Ph.D. Thesis 
(Chemistry Department, IIT, Madras 1997); Kundu notes that 
in solutions the counter ions can affect (renormalise)
the cross over temperature considerably in some cases\\
11) T.K. Sham, Chem. Phys. Lett. {\bf 83} 391 (1981); B. Beagley et
al. J. Phys.: Conens. Matter, {\bf 1} 2395 (1989);
A recent paper,  Ralf Akesson et al., J. Phys. Chem. {\bf 96} 150 (1992),
refers to some modern EXAFS measurements by I. Watanabe and
N. Matsubayashi (Osaka) on the two copper-oxygen distances in
the octahedra.\\
12) B. Bleaney and K.D. Bowers, Proc. Phys. Soc. (London) {\bf A65}
667 (1952)

\end{document}